\documentclass[conference]{IEEEtran}
  \usepackage{cite}

\ifCLASSINFOpdf
   \usepackage[pdftex]{graphicx}
\else
\fi
\usepackage{amsmath}

\usepackage[ruled]{algorithm2e}
\usepackage{booktabs}
\usepackage[normalem]{ulem}
\useunder{\uline}{\ul}{}
\SetKwInOut{Parameters}{Parameters}

\begin{document}
%
\title{Detecting Network Soft-failures with the Network Link Outlier Factor (NLOF)}

\author{\IEEEauthorblockN{Christopher Mendoza}
\IEEEauthorblockA{Department of Electrical and\\Computer Engineering\\
University of Texas at El Paso\\
El Paso, Texas 79968\\
Email: camendoza7@miners.utep.edu}
\and
\IEEEauthorblockN{Venkat Dasari}
\IEEEauthorblockA{U.S. Army Research Laboratory\\
Aberdeen Proving Ground, MD\\
Email: venkateswara.r.dasari.civ@mail.mil}
\and
\IEEEauthorblockN{Michael P. McGarry}
\IEEEauthorblockA{Department of Electrical and \\Computer Engineering\\
University of Texas at El Paso\\
El Paso, Texas 79968\\
Email: mpmcgarry@utep.edu}
}


%


\maketitle

\begin{abstract}
In this paper, we describe and experimentally evaluate the performance of our Network Link Outlier Factor (NLOF) for detecting
soft-failures in communication networks. The NLOF is computed using the throughput values derived from NetFlow records. The flow
throughput values are clustered in two stages, outlier values are determined within each cluster, and the flow outliers are used
to compute the outlier factor or score for each network link. When sampling NetFlow records across the full span of a network, NLOF 
enables the detection of soft-failures across the span of the network; large NLOF scores correlate well with links experiencing 
failure.
\end{abstract}


%
\IEEEpeerreviewmaketitle

\section{Introduction}
\label{sec:intro}
The collection of network data and the application of data analytics (including machine learning) allow the development of technologies to
automate network management. Network management is best characterized using the FCAPS model from the ITU:

\begin{itemize}
  \item \textbf{F}ault detection and correction
  \item \textbf{C}onfiguration and operation
  \item \textbf{A}ccounting and billing
  \item \textbf{P}erformance assessment and optimization
  \item \textbf{S}ecurity assurance and protection
\end{itemize}

In this work we seek to advance the automation of network fault detection. Specifically for network soft-failures that result in diminished
performance. The symptoms of soft-failures are subtle and are therefore difficult to diagnose manually: increased bit errors, occasional 
packet loss, unnecessarily long paths through the network, or congestion control mechanisms unnecessarily reducing throughput. In this work, 
we utilize a suite of data analytics (e.g., clustering and outlier detection) to detect the occurrence of network soft-failures: bit errors,
packet loss. The end result of these data analytics is an outlier score for each network link called the Network Link Outlier Factor (NLOF). 
The NLOF score is an indicator of how likely a link is experiencing a network soft-failure.

\subsection{Related work}
A survey~\cite{DS0516} of recent fault localization techniques expands on the taxonomy presented in \cite{SS0704}. The
taxonomy presented consists of three categories of fault localization: Artificial Intelligence, Model Traversing, and Graph-theoretic.



Much of the related work implements active probing techniques \cite{MMHTOM0515, D1206, CW1016}. These techniques use probe 
messages to infer the state of links and require optimal probe placement~\cite{NSL0308, CQMQB0310, NS1107} to trade off measurement with 
resource consumption. More recent work~\cite{MTG0818} uses passive data (e.g., number of: flows, lost packets, average packet delay) 
and compares the performance of several machine learning techniques (e.g., random forests and multi-layer perceptrons) to localize faults. 
Their results are compared to the active probing technique in \cite{CW1016}. Other recent work in the domain of optical networking uses
passively collected physical layer data and machine learning to detect and/or localize faults~\cite{VSRCCLPCCYV0118, SMCT0318}.

Some related work uses a hybrid-approach~\cite{TAB0505, TAB0308} that mixes active probing with passive data collection. In \cite{TAB0505} 
the authors present a fault localization framework named Active Integrated fault Reasoning (AIR). AIR uses passive monitoring to compile 
a set of observed symptoms. The framework then generates sets of faults that may be causing the symptoms. Each of these fault sets is 
tested to validate if any of them are credible. If none are credible then there is likely to be a symptom that was not observed by the 
passive monitoring. To identify if the likely symptom is present active probes are 
used. Afterward, the sets of faults go through the credibility test again.

Software Defined Networks~\cite{AAK0914, SSCPD1011} including OpenFlow~\cite{MABPPRST0408} along with advances in machine learning are 
sparking a resurgence in network fault localization. Software Defined Networking allows for a broader view of the network, providing a
simpler way to obtain network topology information~\cite{PPTI0316} for fault localization.

As far as we know, our work is the first to use passive NetFlow and topology data to detect/localize network faults.

\subsection{Outline}
In Section \ref{sec:NLOF} we describe our suite of data analytics resulting in the NLOF score for each network link. In Section 
\ref{sec:experiments} we describe our NS-3 experiments to evaluate the failure detection performance of NLOF and in Section \ref{sec:results}
we present and discuss the results of those experiments. Finally, we discuss our conclusions and outline paths for future work in Section
\ref{sec:conclusion}.

\section{Detecting Network Link Soft Failures}
\label{sec:NLOF}
We propose a method to detect network link soft-failures using NetFlow data. Specifically, we use the average throughput of flows from
their collected NetFlow records. If the collected data consists of flows traversing the full span of the network, we believe it will allow
us to detect any soft-failures in the full topology. We propose using flow throughput outlier detection to assist with the detection of 
network link soft-failures. Using topology and routing information we can correlate flows with the network links they traverse. We 
hypothesize that a network link experiencing a soft-failure will cause the flows traversing that link to exhibit abnormal throughput. 
Therefore, a network link carrying many flows that are throughput outliers is one that is experiencing a soft-failure.

\begin{figure*}[!ht]
\centering
\includegraphics[scale = 0.39]{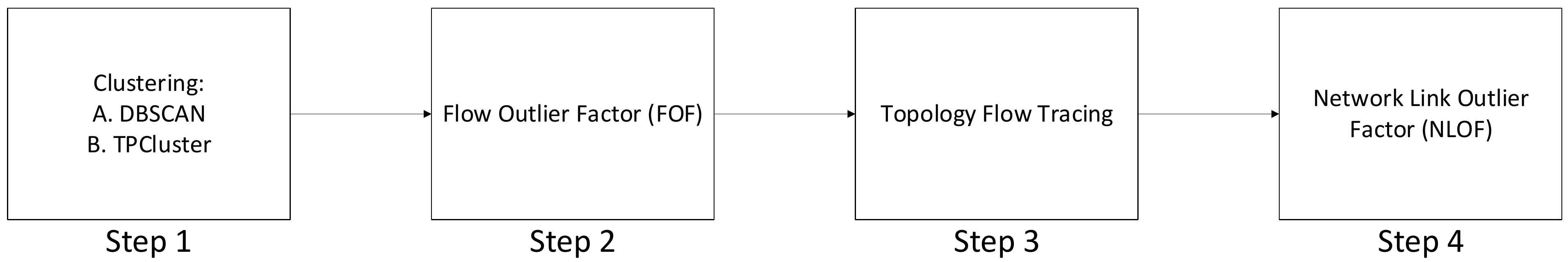}
\caption{Data analytics pipeline to compute Network Link Outlier Factor (NLOF)}
\label{fig:pipeline}
\end{figure*}

Using outlier detection directly on the average throughput of all flows requires the assumption that a majority of flows do
not have abnormal throughput. Since this may not be a reasonable assumption, we first cluster the throughput of flows into the set
of clusters we believe will naturally exist in a network and then identify the outliers within those throughput clusters. Our full 
technique to detect network link soft-failures consists of: 1) flow throughput clustering, 2) flow throughput outlier detection using 
an outlier score, 3) tracing flows on the network topology using routing information, and 4) network link outlier score computation from 
flow outlier scores. Figure \ref{fig:pipeline} illustrates this 4-stage analytics pipeline for detecting network link soft-failures.

\subsection{Flow Throughput Clustering (DBSCAN and TPCluster)}
Network link soft-failures could cause a majority of flows to exhibit reduced throughput. As a result, we cannot immediately apply
outlier detection techniques to the average flow throughput values. We propose to first organize average flow throughput into
clusters. Network topologies will generally employ several network link transmission rates and the average flow throughput values
will be limited by those values. In isolation, the average throughput of a flow will be limited by the bottleneck network link that
it traverses. Let $\alpha$ be either the original generated throughput (or bitrate) of the flow or the bitrate at which the flow
enters the network of interest, $\Gamma(i)$ be the network link rate of the $i$th network link a flow traverses in the network of
interest. Then, in isolation, the average throughput of that flow will be:

\begin{equation}
\min \left\{\alpha, \min_{\forall_i} \left\{\Gamma(i)\right\}\right\}
\end{equation}

Flows will often share network links with other flows. Let $n(i)$ be the number of flows sharing the $i$th network link a flow traverses. 
Let's assume that flows share network links equally. Then, while sharing network links with other flows, the
average throughput of that flow will be:

\begin{equation}
\min \left\{\alpha, \min_{\forall_i} \left\{\frac{\Gamma(i)}{n(i)}\right\}\right\} \mbox{.}
\end{equation}

%
%

%
%
%
%
%

Suppose we had a network topology with two different link rates, 1 Gbps and 100 Mbps and a network link was never shared by more than
4 flows at a time. In this case we would have 8 different values for average flow throughput in descending order (1 Gbps, 500 Mbps, 333 Mbps,
250 Mbps, 100 Mbps, 50 Mbps, 33.3 Mbps, and 25 Mbps). Network links experiencing soft-failures will reduce these average flow throughput 
values for flows traversing those links. Therefore, any average flow throughput values deviating significantly from these 8 discrete values
are likely affected by a network soft-failure.



Since we generally do not know the set of network link transmission rates nor the number of flows sharing network links, we let unsupervised
machine learning (specifically, clustering) find these values for us. We select Density-Based Spatial Clustering of Applications with Noise 
(DBSCAN)\cite{EKSX0896} as our first stage of clustering since it naturally finds dense and potentially non-convex clusters without knowing 
the number of clusters a-priori. DBSCAN produces a number of clusters as well as a set of data points labeled as ``noise'' that do not fit 
into any of the clusters.

A second stage of clustering is performed since the flows affected by soft-failures may begin to form their own dense cluster. In this 
second stage that we call TPCluster, we want to combine adjacent clusters if they are within a proximity to each other that suggests one
may be a performance degraded set of the other. TPCluster provides a dynamic range based on the throughput context of the DBSCAN clusters. 
TPCluster uses two parameters, throughput ratio (\textit{tpr}) and throughput deviation (\textit{tpdev}). \textit{tpr} should be set to the 
maximum reasonable performance degradation of a throughput class and \textit{tpdev} should be set to the deviation that you might expect to 
see from a throughput class, to help cluster the flows labeled as ``noise'' into the appropriate cluster. Algorithm \ref{Alg:TPCluster} 
shows how TPClusters are formed.

\begin{algorithm}[!ht]
\label{Alg:TPCluster}
\SetAlgoLined
\KwData{Set of DBSCAN Clusters and the DBSCAN Noise Cluster}
\Parameters{TPRatio (tpr), Throughput Deviation (tpdev)}
\KwResult{Set of TPClusters}
Sort DBSCAN Clusters In Descending Throughput Order\;
j = 0\;
\For{cluster[i] in DBSCAN clusters}{
  \If{cluster[i] has not been combined}{
      TPCluster[j] = aggregation of DBSCAN Clusters within range of ((1 - tpr)*$cluster[i]_{max}$, $cluster[i]_{max}$)\;
      j++\;
  }
}

\For{flow[j] in the DBSCAN noise cluster}{
   k = 0\;
   \For{$cluster[i]$ in TPClusters}
   {
       dist = $cluster[i]_{max} - flow[j]_{TP}$\;
       \If{$dist \geq -tpdev * TPCluster[i]_{max}$}
       {
           $candidate[j][k]_{dist}$ = dist\;
           $candidate[j][k]_{cluster}$ = $cluster[i]$\;
           k++\;
       }
   \eIf{$k > 0$}{
       ind = $\underset{k}{\operatorname*{argmin}} \{candidate[j][k]_{dist} \}$\;
       Add $flow[j]$ to $candidate[j][ind]_{cluster}$\;
       }{
       Add $flow[j]$ to TPCluster with largest throughput\;
       }
   }
}
\caption{TPCluster}
\end{algorithm}

\subsection{Flow Outlier Factor (FOF)}
Now that TPClusters have been defined we must now choose a point in each cluster to be the representative "normal" point 
($Cluster_{normal}$) i.e. the point with a reasonable desirable performance. A method is to just use the point in the cluster with 
the highest performance however, we propose to use k-means clustering to use the cluster center as the representative point and as 
k increases the more aggressive the representative point will be. Algorithm \ref{Alg:FOF} shows how the Flow Outlier Factor (FOF) of 
each flow is computed.

\begin{algorithm}[!ht]
\label{Alg:FOF}
\SetAlgoLined
\KwData{Set of TPClusters and their constituent flows}
\KwResult{FOF Scores for each flow}
\For{cluster[i] in TPClusters}{
    Separate cluster[i] into k clusters using K-Means\;
    $cluster[i]_{normal}$ = mean of K-Means cluster with the highest throughput\;
    \For{flow[i][j] in cluster[i]}{
        $flow[i][j]_{FOF}$ = $\frac{cluster[i]_{normal} - flow[i][j]_{TP}}{cluster[i]_{normal}}$
    }
}
\caption{Compute FOF}
\end{algorithm}


\subsection{Topology Flow Tracing}
In this step we associate network links with the flows that traverse them. To make this association, flows are traced on the network 
topology using routing information. We use the NetworkX Python package to trace flows on the topology of a network assuming shortest
path routing.


\subsection{Network Link Outlier Factor (NLOF)}
In this final step, we compute the outlier score for each network link (i.e., the NLOF). The flow outlier scores (i.e., FOFs) are
used to compute the NLOF for the network links the flows traverse. Outlier flows are determined by a threshold on their FOF and 
the NLOF is computed to be the ratio of outlier flows to total flows traversing the network link. The NLOF is computed using 
Algorithm \ref{Alg:NLOF}.

\begin{algorithm}[!ht]
\label{Alg:NLOF}
\SetAlgoLined
\KwData{Set of Links with their associated flows}
\Parameters{Outlier threshold}
\KwResult{NLOF score for each Link}
\For{$Link[i]$ in Links}{
    outlierFlows = 0\;
    \For{Each $flow[i][j]$ in $Link[i]$}{
        \If{$flow[i][j]_{FOF} >$ Threshold}{
            outlierFlows++\;
        }
    }
    $Link[i]_{NLOF}$ = $\frac{outlierFlows}{Flows \ in \ Link[i]}$
}
\caption{NLOF Computation}
\end{algorithm}

\section{Experiments}
\label{sec:experiments}
\begin{figure*}[!t]
\centering
\begin{tabular}{cc}
\includegraphics[width= 0.47\textwidth]{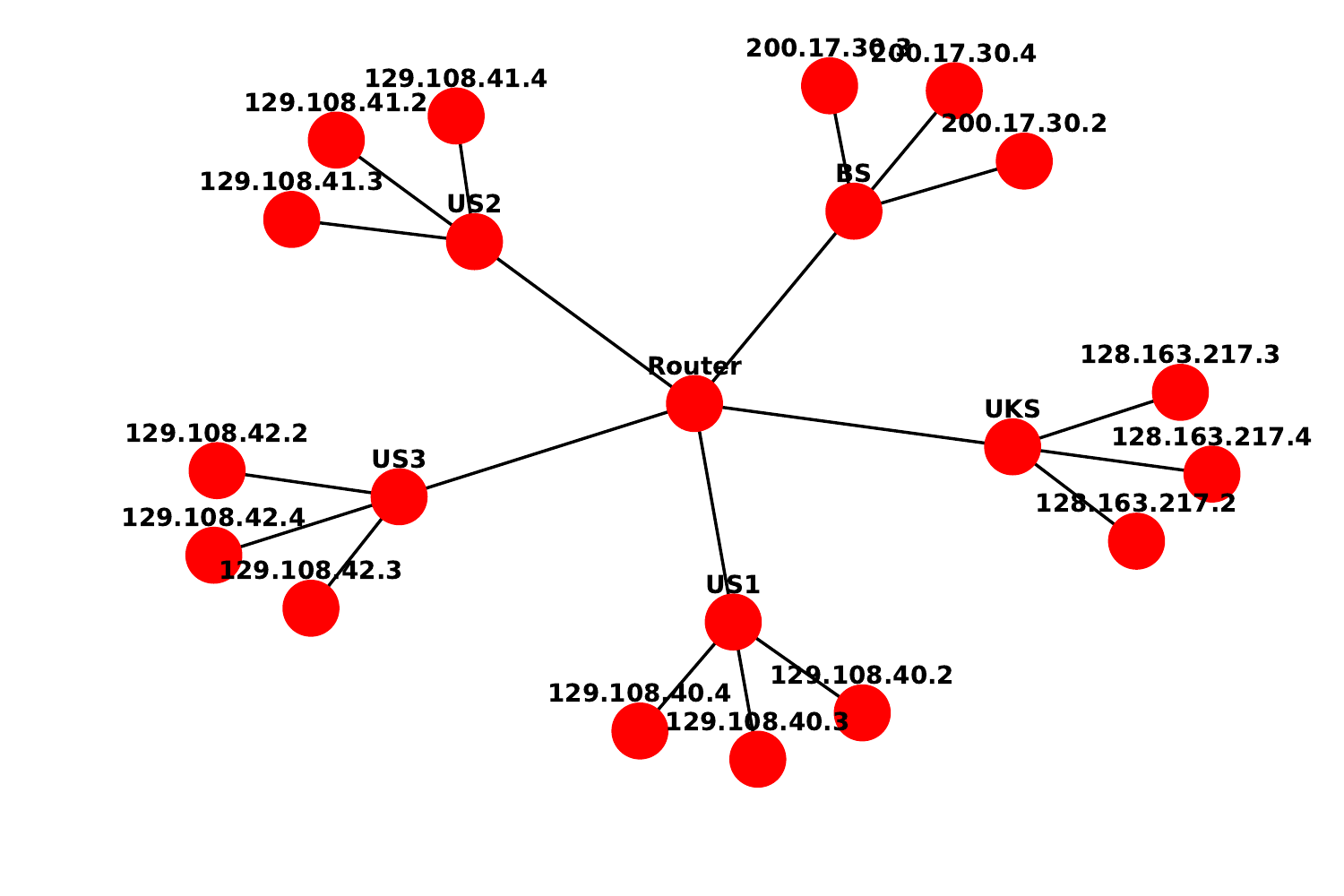}&
\includegraphics[width= 0.47\textwidth]{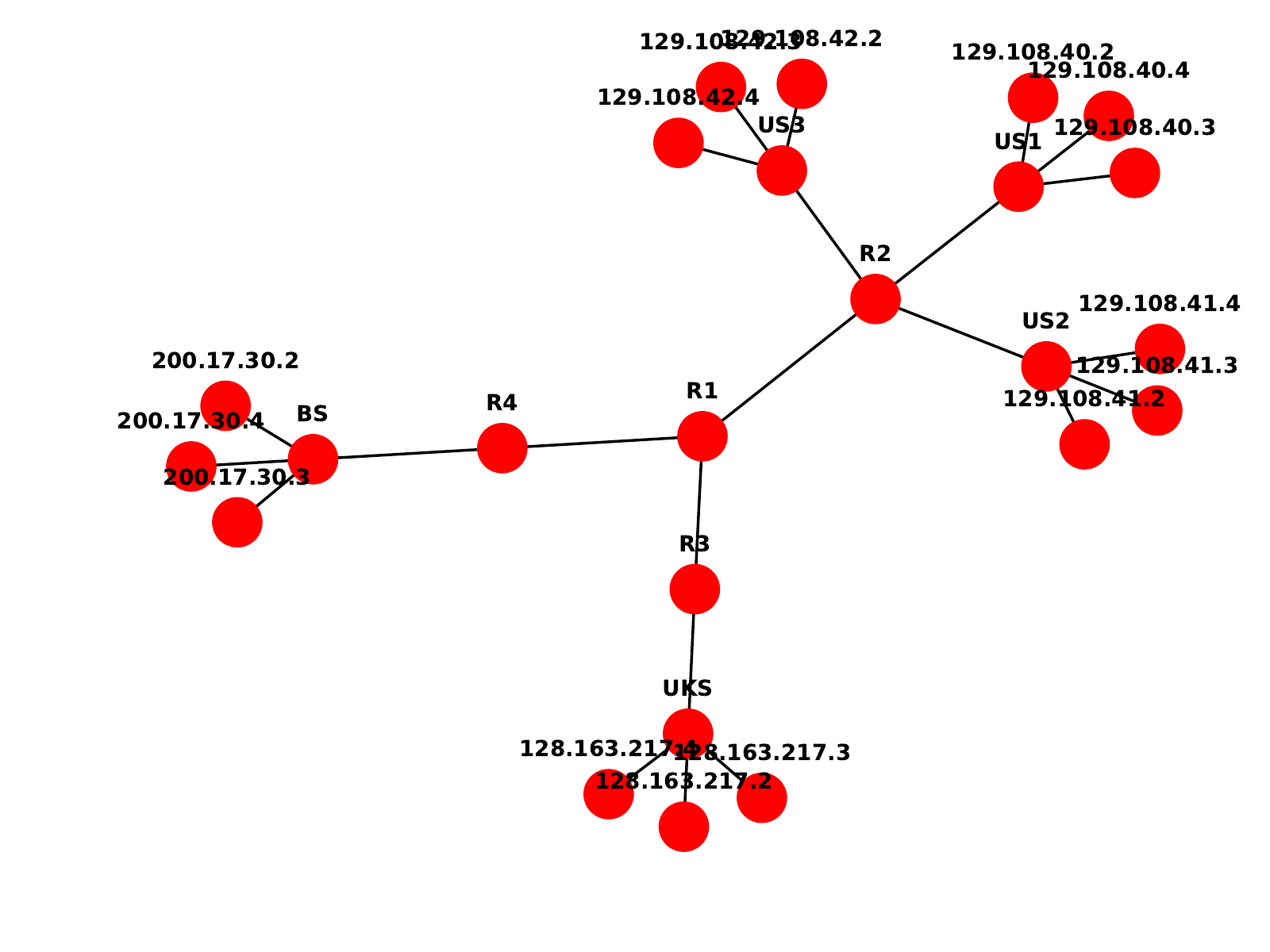}\\
a) Topology 1&
b) Topology 2\\
\end{tabular}
\caption{NS-3 Topologies}
\label{fig:Topologies}
\end{figure*}

To evaluate the performance of NLOF, we utilize NS-3 simulation experiments. Figure \ref{fig:Topologies} show the two 
topologies we simulated. For each topology 3 experiments were run, for a total of 6 experiments. For each experiment all links were set 
to have a data rate of 10 Mbps. US1, US2, US3, UKS and BS are all OpenFlow switches implemented with the OpenFlow 1.3 module. The nodes 
and routers populate their routing tables using Routing Information Protocol (RIP). During each simulation 5000 On/Off flows were produced 
at a rate of either 1Mbps or 1Kbps between two randomly selected hosts in the network, the only exception is test 6, it had 2 additional 
throughput classes which are 10 Kbps and 2 Mbps for a total of 4. To collect the data from the simulation the built-in flow monitor model 
library was used. The probes were installed on all nodes to capture all the traffic in the network. Table \ref{Tab:Parameters} shows the 
configuration of the simulation for each test.

The flow monitor library outputs the files in XML format, we then parsed the XML file to construct a pandas DataFrame to resemble flow 
records. The produced DataFrame will be in an acceptable format for SciKitLearn's DBSCAN clustering method. The DBSCAN clustering was 
done using the parameters eps = 100 and min\_samples = 50 which produced clusters that could then be combined to form TPClusters. TPClusters 
were formed using Algorithm \ref{Alg:TPCluster} with parameter values of tpr = 0.3, tpdev = 0.1 and k = 2. The flows were traced to put 
each flow into every network link that it traversed, assuming that the flow will take the shortest path which can be obtained using the 
NetworkX shortest path function. Finally the NLOF for each object was calculated using Algorithm \ref{Alg:NLOF} with an FOF threshold value 
of 0.1.

\begin{table}[!ht]
\centering
\caption{Test Parameters}
\label{Tab:Parameters}
\resizebox{\columnwidth}{!}{
\begin{tabular}{|c|c|c|c|c|}
\hline
\textbf{Test} & \textbf{Links with Errors} & \textbf{Error Rate} & \textbf{Topology} & \textbf{Throughput Classes} \\ \hline
1 & None & 0 & 1 & 100 Kbps, 1Mbps \\ \hline
2 & (’129.108.40.2’, ’US1’) & 0.1 & 1 & 100 Kbps, 1Mbps \\ \hline
3 & \begin{tabular}[c]{@{}c@{}}(’129.108.42.4’, ’US3’)\\ (’129.108.41.3’, ’US2’)\\ (’129.108.40.2’, ’US1’)\end{tabular} & \begin{tabular}[c]{@{}c@{}}0.1\\ 0.1\\ 0.1\end{tabular} & 1 & 100 Kbps, 1Mbps \\ \hline
4 & None & 0 & 2 & 100 Kbps, 1Mbps \\ \hline
5 & (’BS’, ’R4’) & 0.1 & 2 & 100 Kbps, 1Mbps \\ \hline
6 & \begin{tabular}[c]{@{}c@{}}(’129.108.40.2’, ’US1’)\\ (’128.163.217.2’, ’UKS’)\\ (’129.108.42.4’, ’US3’)\end{tabular} & \begin{tabular}[c]{@{}c@{}}0.05\\ 0.03\\ 0.01\end{tabular} & 2 & \begin{tabular}[c]{@{}c@{}}10 Kbps, 100 Kbps, \\ 1 Mbps, 2Mbps\end{tabular} \\ \hline
\end{tabular}
}
\end{table}

\section{Results}
\label{sec:results}
\begin{figure*}[!t]
\centering
\begin{tabular}{cc}
\includegraphics[width= 0.7\columnwidth]{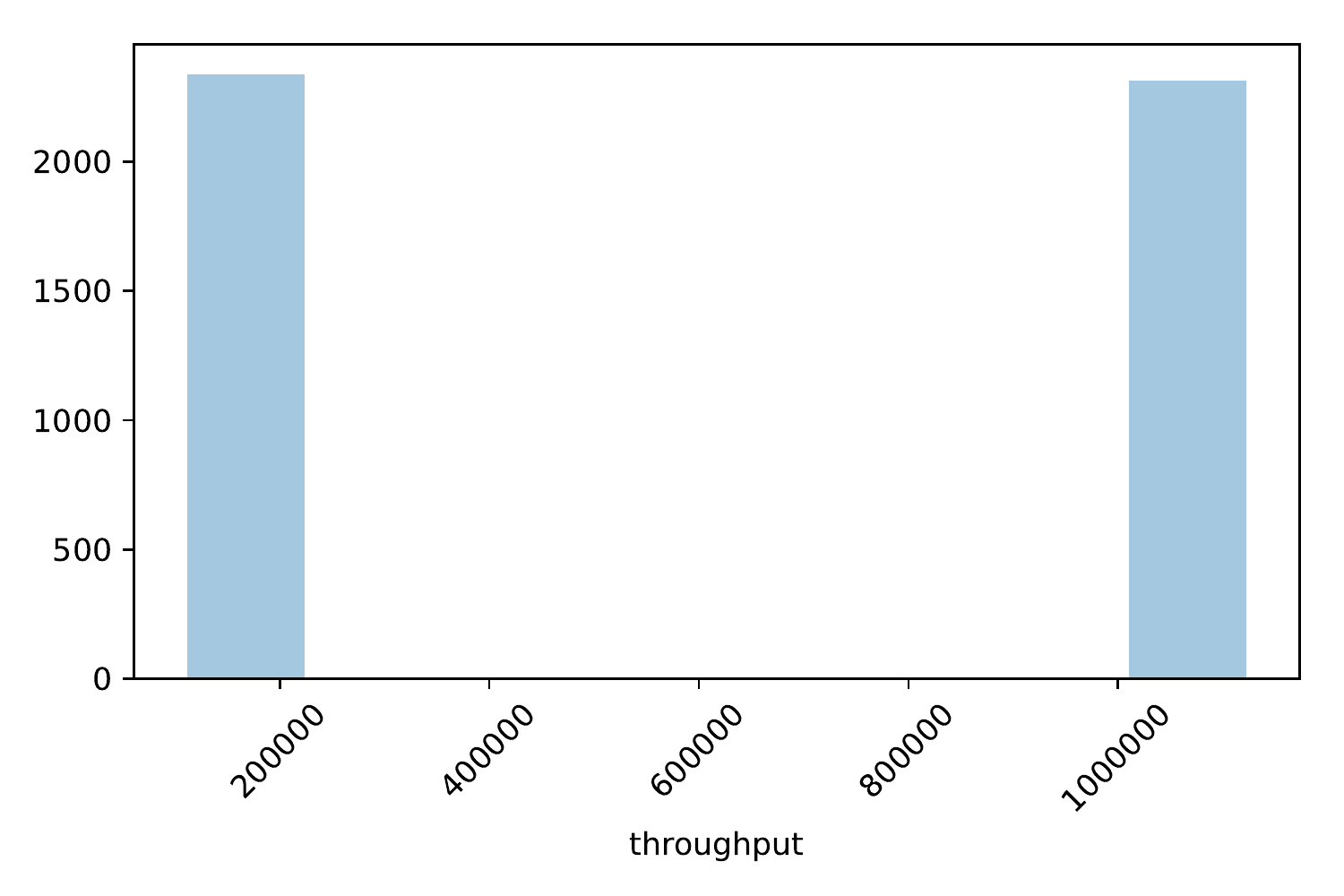}&
\includegraphics[width= 0.7\columnwidth]{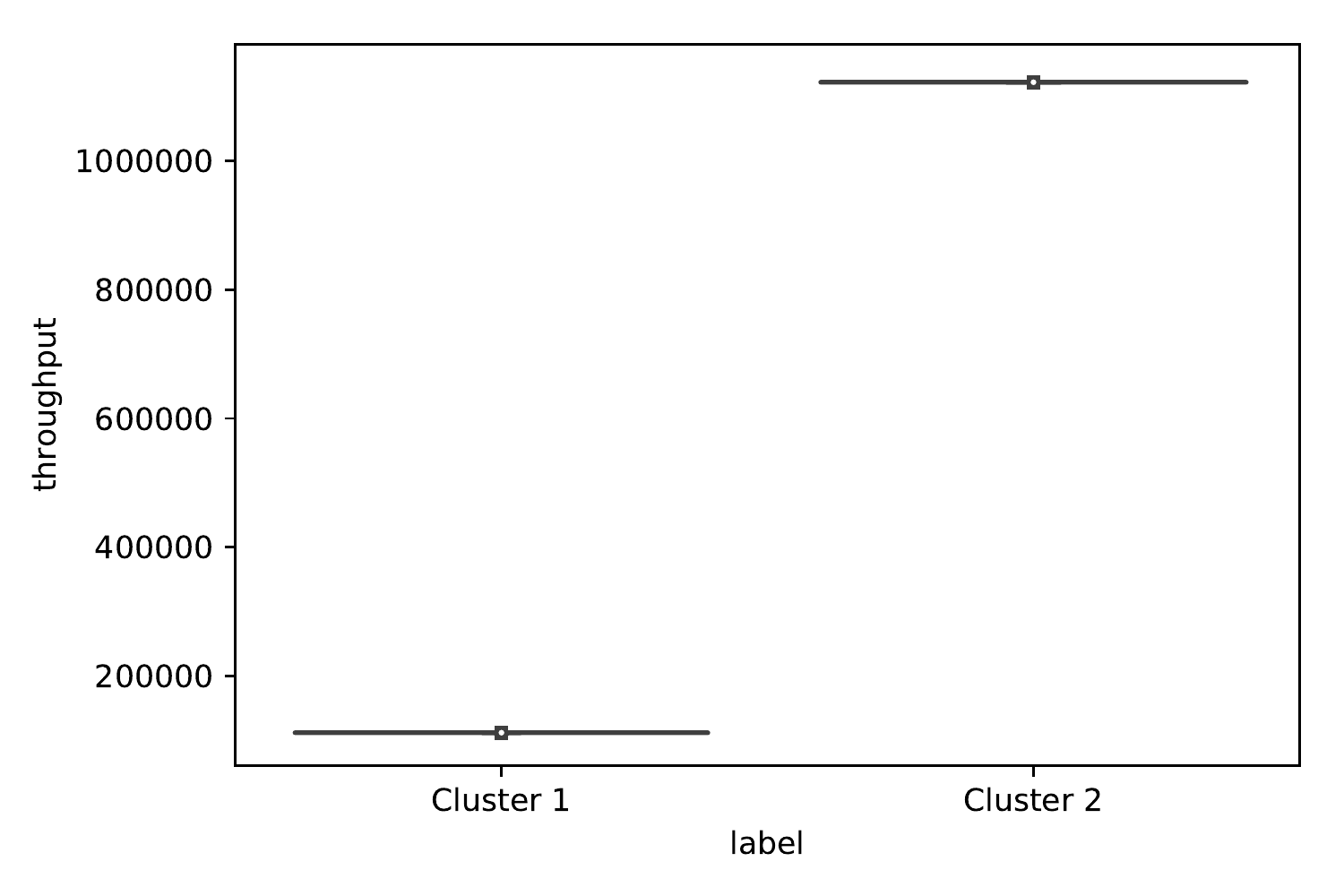}\\
a) Input Data Distribution&
b) Violin Plot of TPClusters\\
\end{tabular}
\caption{Input data and corresponding TPClusters for test 1}
\label{fig:ClustT1}
\end{figure*}

\begin{figure*}[!t]
\centering
\begin{tabular}{cc}
\includegraphics[width= 0.7\columnwidth]{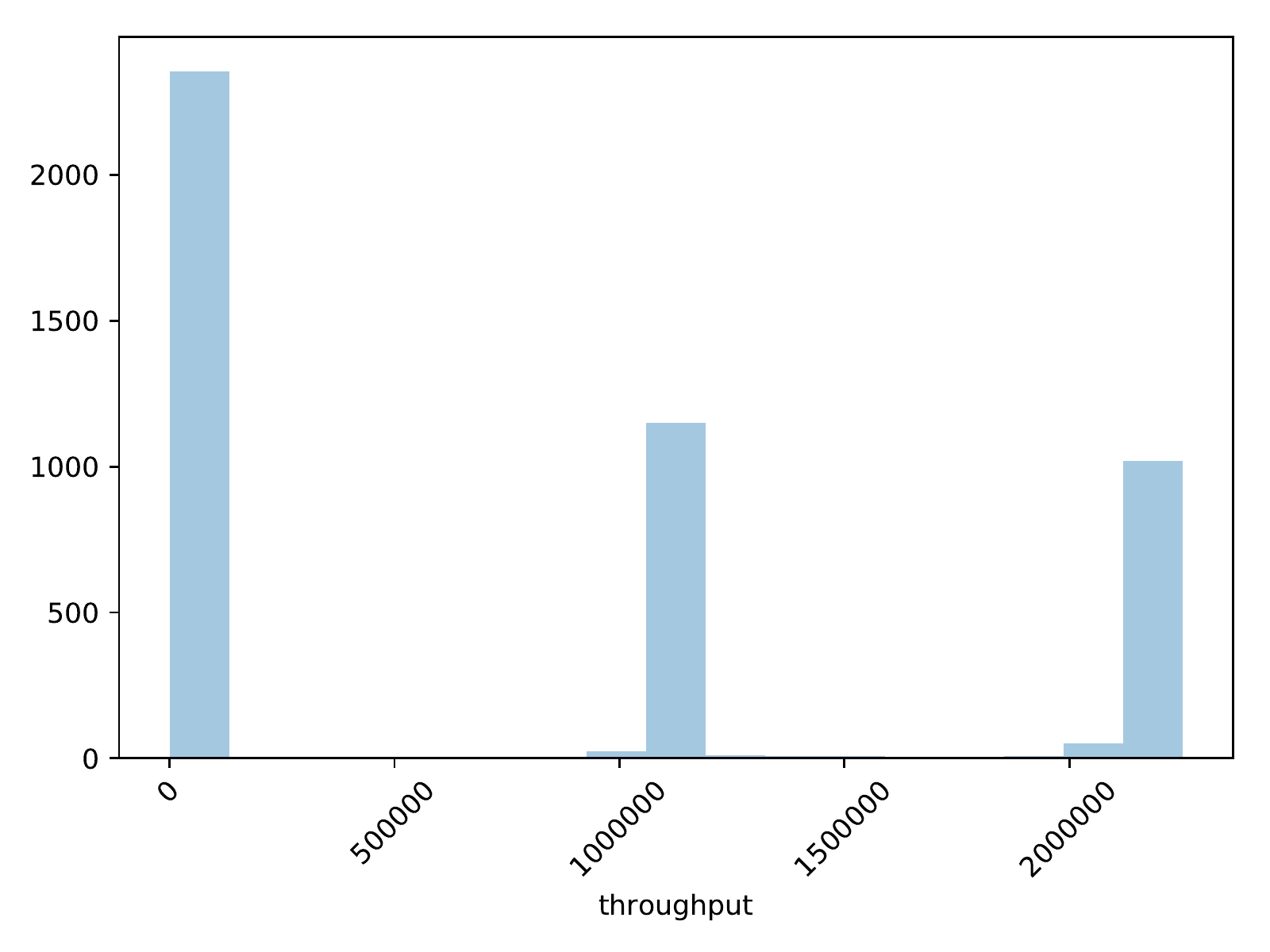}&
\includegraphics[width= 0.7\columnwidth]{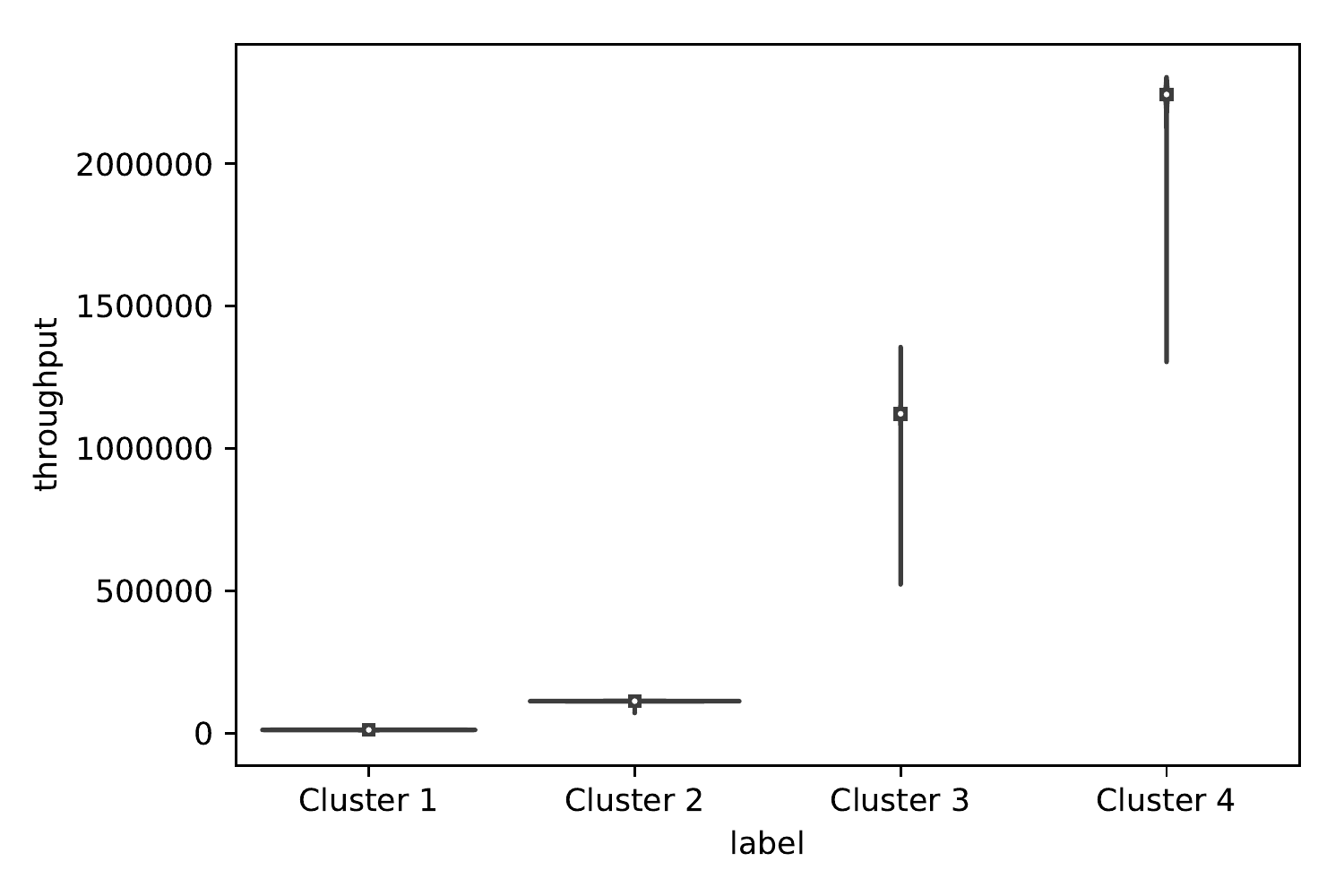}\\
a) Input Data Distribution&
b) Violin Plot of TPClusters\\
\end{tabular}
\caption{Input data and corresponding TPClusters for test 6}
\label{fig:ClustT2}
\end{figure*}

\begin{figure*}[!t]
\centering
\begin{tabular}{cc}
\includegraphics[width= 0.6\columnwidth]{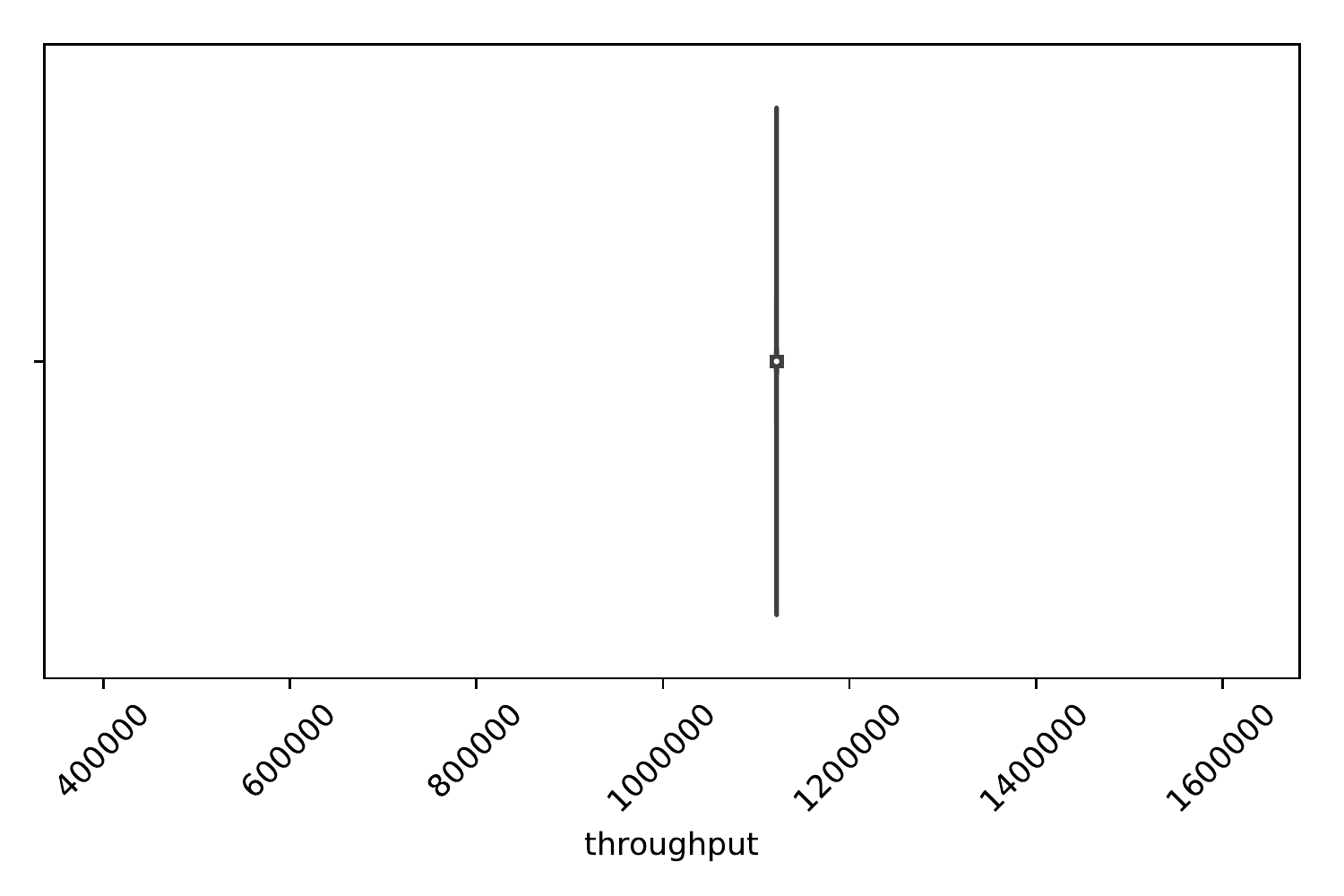}&
\includegraphics[width= 0.6\columnwidth]{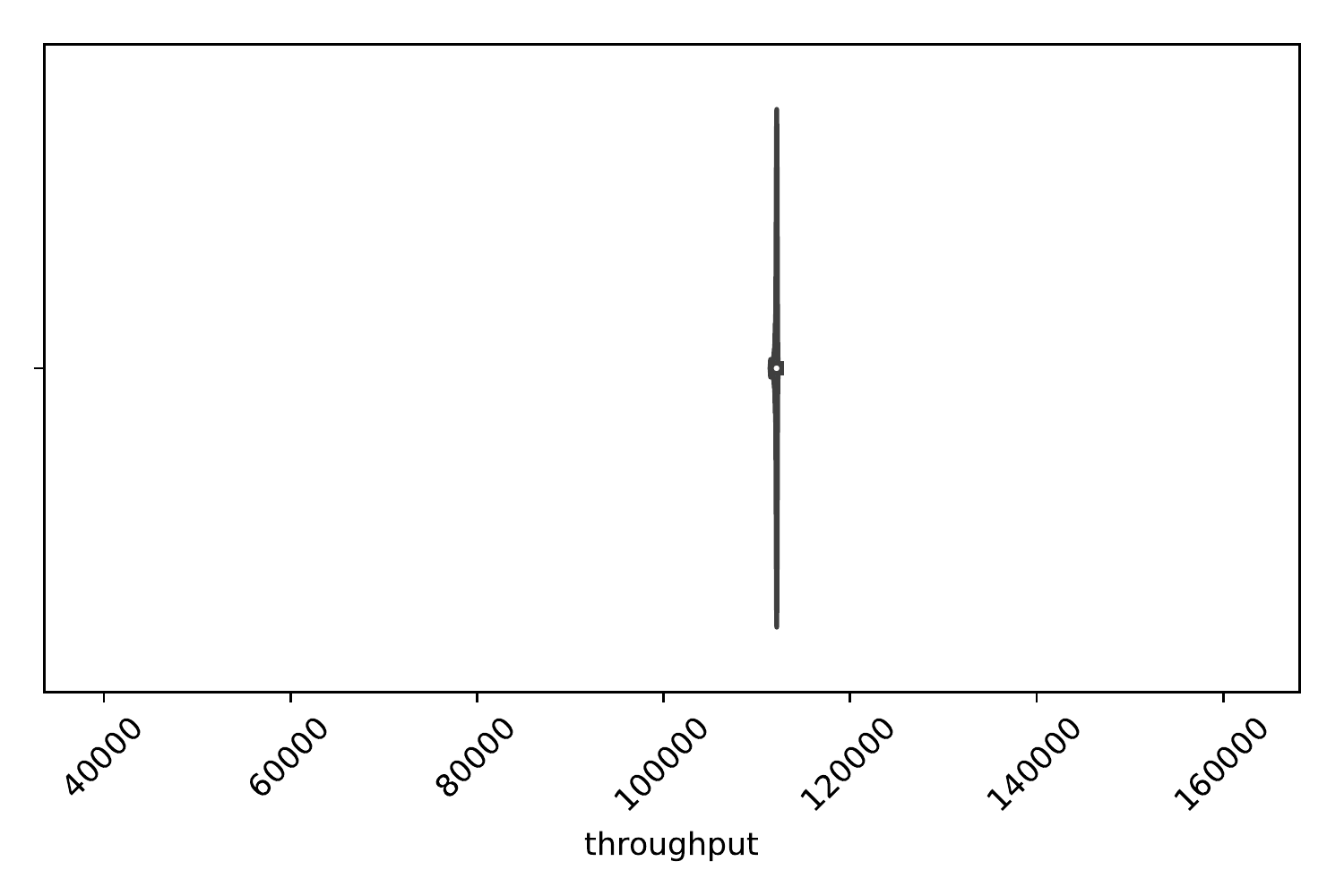}\\
a) Cluster 1 Distribution&
b) Cluster 2 Distribution\\
\end{tabular}
\caption{Distribution of test 1 clusters}
\label{fig:TP1}
\end{figure*}

\begin{figure*}[!t]
\centering
\begin{tabular}{cccc}
\includegraphics[width= 0.23\textwidth]{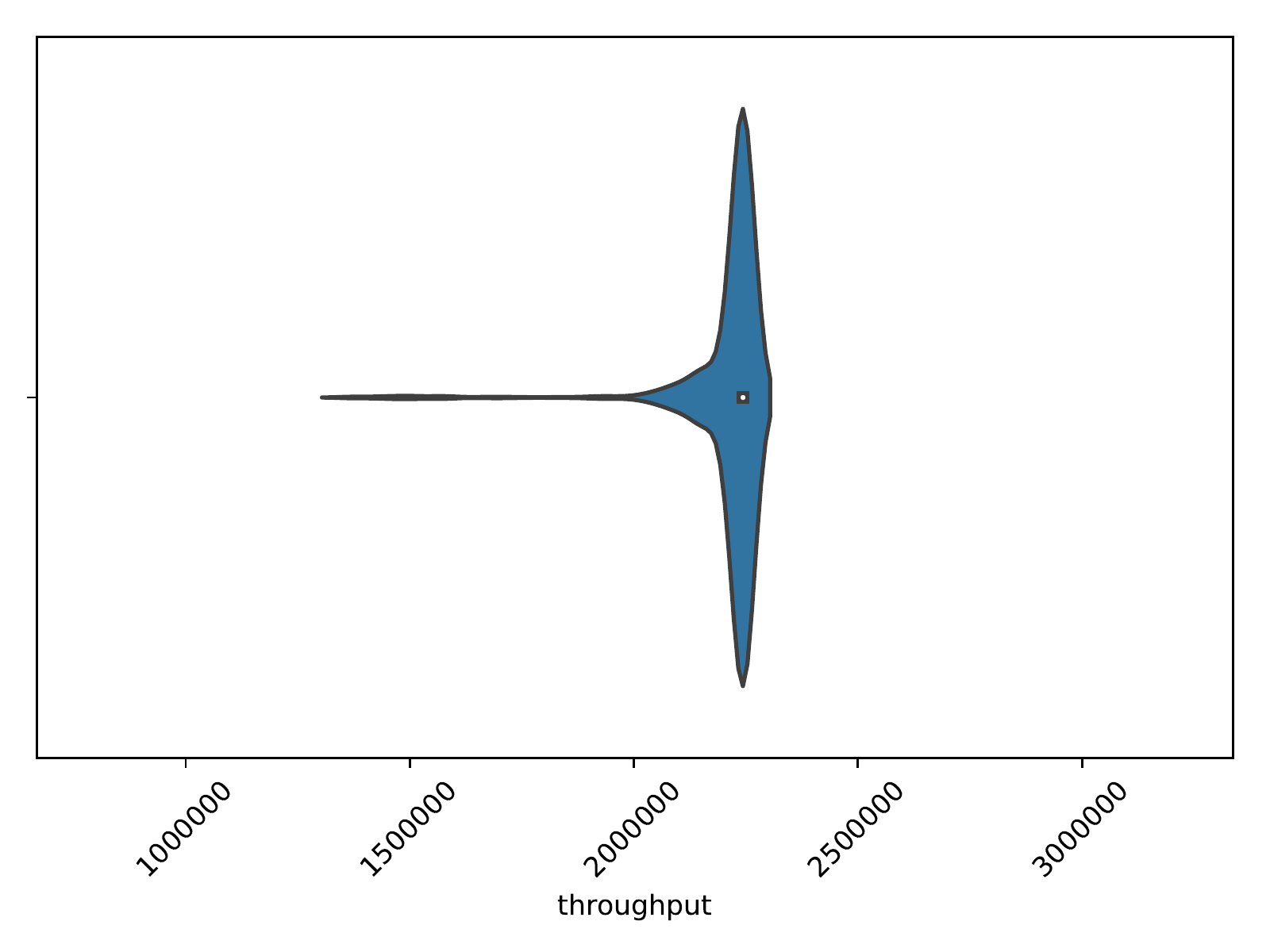}&
\includegraphics[width= 0.23\textwidth]{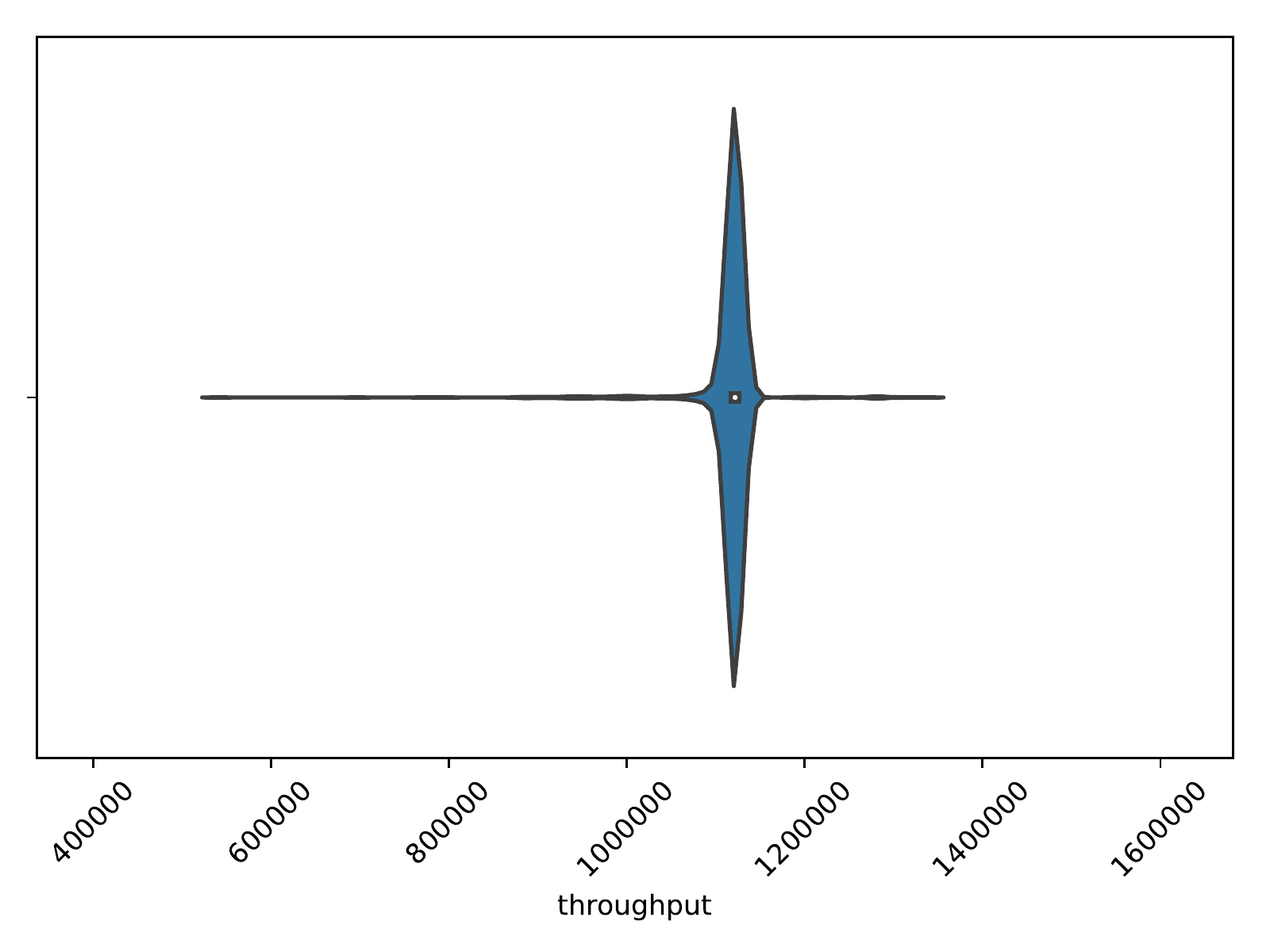}&
\includegraphics[width= 0.23\textwidth]{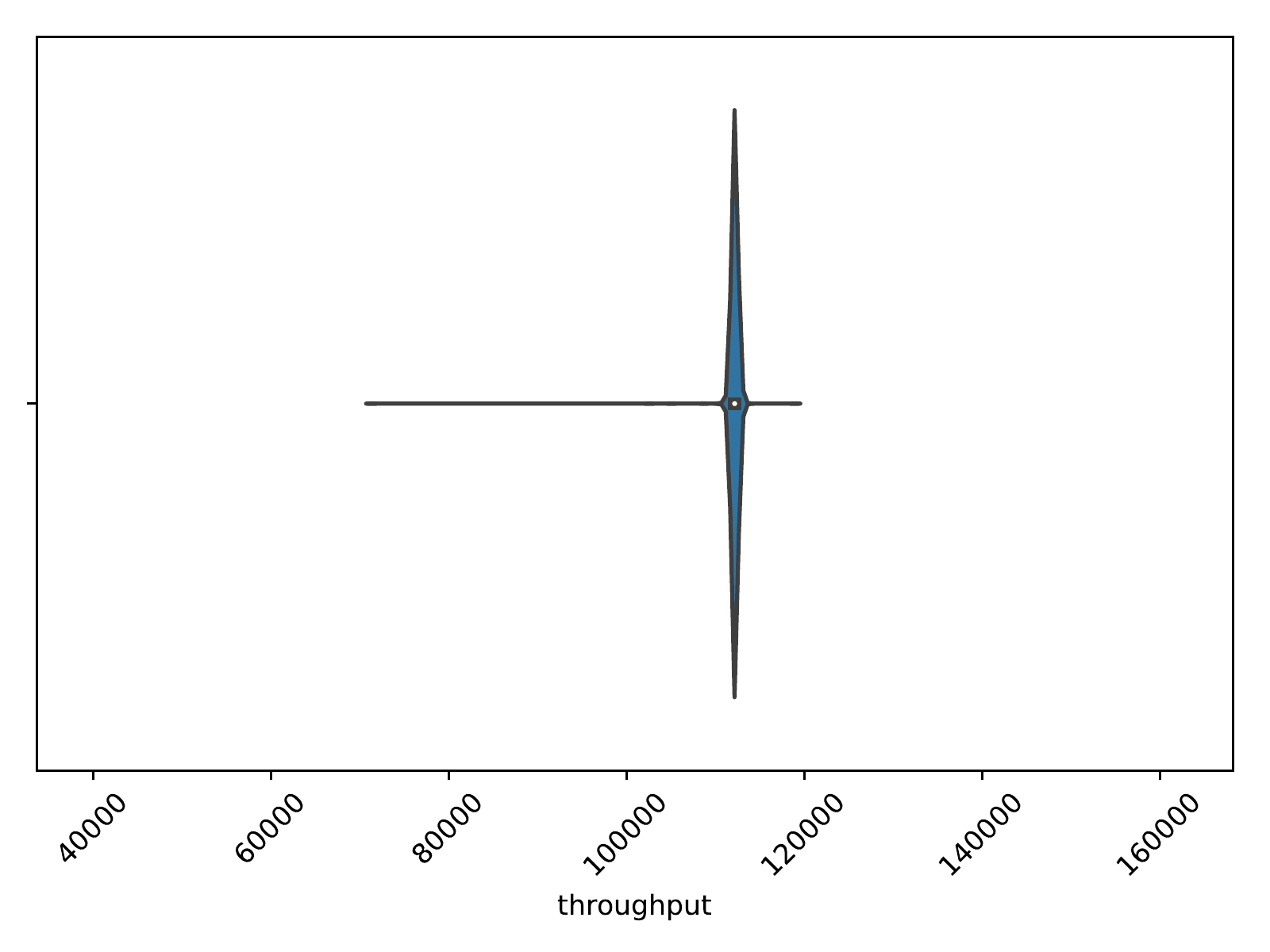}&
\includegraphics[width= 0.23\textwidth]{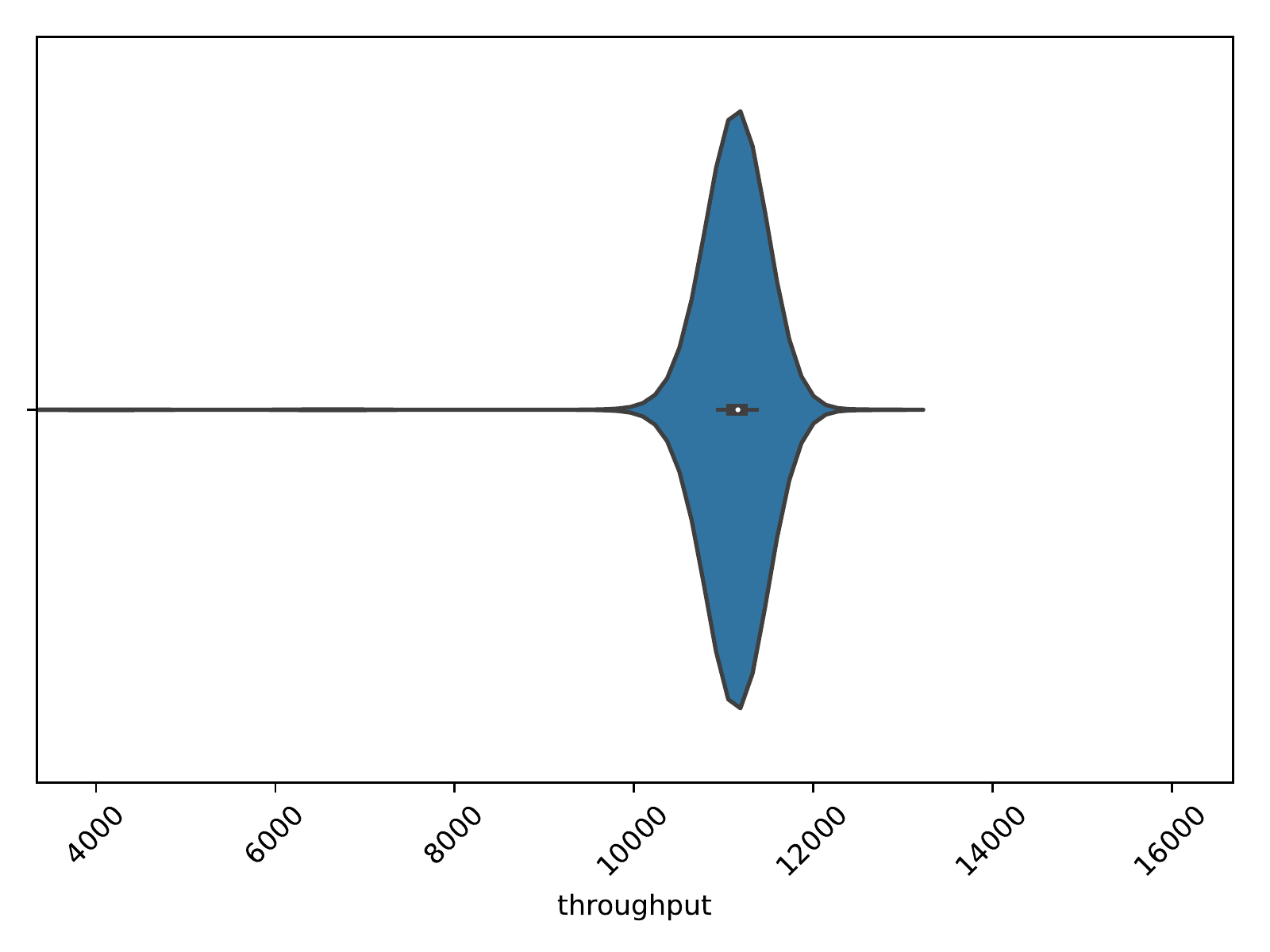}\\
a) Cluster 1 Distribution&
b) Cluster 2 Distribution&
c) Cluster 3 Distribution&
d) Cluster 4 Distribution\\
\end{tabular}
\caption{Distribution of test 6 clusters}
\label{fig:TP2}
\end{figure*}

Figures \ref{fig:ClustT1} and \ref{fig:ClustT2} show the flow throughput distribution (left-side sub-plot) and the corresponding TPClusters 
in a violin plot (right-side sub-plot) for tests 1 and 6 respectively. Figures \ref{fig:TP1} and \ref{fig:TP2} show the flow throughput 
distributions of each TPCluster for tests 1 and 6 respectively. As shown in Figures \ref{fig:ClustT1} and \ref{fig:ClustT2}, the TPClusters 
formed as expected i.e. one cluster for every throughput class. For test 1 we have the two throughput classes 100 Kbps and 1 Mbps with 2 
corresponding clusters. For test 6 we have 4 TPClusters one for each of the throughput classes. More importantly the points labeled as noise 
by DBSCAN are moved into their appropriate TPCluster. For test 1 the cluster distributions have a small range that clearly indicates none of 
the flows have poor throughput performance within the context of their cluster. Figure \ref{fig:TP2} shows more interesting flow throughput 
distributions, this time there are four separate clusters, which all have flows farther away from the $Cluster_{normal}$ which can be seen 
visually by the larger range of each cluster. $Cluster_{normal}$ in this case is located in the upper half of the cluster distribution. The 
large range of the clusters indicates that there are flows with poor throughput performance belonging to these clusters. Our two-step 
clustering organizes the flows to properly identify those experiencing poor throughput performance.

Tables \ref{Tab:Topo1Results} and \ref{Tab:Topo2Results} show the results obtained for each test for topology 1 and 2 respectively. 
The links and NLOF scores that are in bold are the links that were set to have errors in the simulation, which correspond to 
Table \ref{Tab:Parameters}. The tables show that for Tests 1 and 4, both of which have no poor performing links, there is a NLOF 
score of 0 for all links. Test 2 has one link with a packet error rate of 0.1, that link has the highest NLOF score by a wide margin. 
Test 5 also has one link with a packet error rate of 0.1. However, this time it is a link connected between two nodes with high 
centrality rather than a link near the edge of the network. A large portion of the network traffic will go through this errored link, 
which explains the much higher NLOF scores in general compared to Test 2. Something interesting to note in Test 5 is that the link with 
a non-zero error rate does not have the highest NLOF score. This is likely caused by the fact that all the traffic to or from 
node "200.17.30.4" must go through the link with a non-zero error rate if communicating with a node not connected to the "BS" switch. 
One last thing to note about Test 5 is that the edges (BS,R4) and (R1,R4) have an identical NLOF score, which is easily explained since 
all the traffic that goes through one of those links must go through the other. Test 3 shows a different scenario now with 3 separate 
links all having a packet error rate of 0.1. Due to the fact that there are more links with non-zero error rates the NLOF scores will 
have larger values since there will be more poor performing traffic. Even in this scenario the NLOF score gives an idea as to which 
links are the ones with non-zero error rates, as the 3 of the top 4 NLOF scores are the links we are looking for as shown in bold in 
Table \ref{Tab:Topo1Results}. For Test 6 the packet error rates were lowered by a significant amount and the 3 links with non-zero 
errors are all different. Test 6 also had the added change of 2 extra throughput classes. The results are as expected, the 3 links 
with the highest NLOF scores are the 3 errored links.

\begin{table*}[!ht]
\centering
\caption{Topology 1 Experiment Results}
\label{Tab:Topo1Results}
\begin{tabular}{@{}cc | cc | cc@{}}
\toprule
\multicolumn{2}{c}{\textbf{Test 1}} & \multicolumn{2}{c}{\textbf{Test 2}} & \multicolumn{2}{c}{\textbf{Test 3}} \\ \midrule
{\ul \textbf{Link}} & {\ul \textbf{NLOF}} & {\ul \textbf{Link}} & {\ul \textbf{NLOF}} & {\ul \textbf{Link}} & {\ul \textbf{NLOF}} \\
('128.163.217.2', 'UKS') & 0 & \textbf{('129.108.40.2', 'US1')} & \textbf{0.084057971} & \textbf{('129.108.42.4', 'US3')} & \textbf{0.462962963} \\
('128.163.217.3', 'UKS') & 0 & ('US1', 'Router') & 0.014066496 & \textbf{('129.108.41.3', 'US2')} & \textbf{0.324675325} \\
('128.163.217.4', 'UKS') & 0 & ('129.108.40.4', 'US1') & 0.005957447 & ('US3', 'Router') & 0.293251534 \\
('129.108.40.2', 'US1') & 0 & ('200.17.30.4', 'BS') & 0.005957447 & \textbf{('129.108.40.2', 'US1')} & \textbf{0.259668508} \\
('129.108.40.3', 'US1') & 0 & ('129.108.42.4', 'US3') & 0.005076142 & ('129.108.40.3', 'US1') & 0.12716763 \\ \bottomrule
\end{tabular}
\end{table*}

\begin{table*}[]
\centering
\caption{Toplogy 2 Experiment Results}
\label{Tab:Topo2Results}
\begin{tabular}{@{}cc | cc | cc@{}}
\toprule
\multicolumn{2}{c}{\textbf{Test 4}} & \multicolumn{2}{c}{\textbf{Test 5}} & \multicolumn{2}{c}{\textbf{Test 6}} \\ \midrule
{\ul \textbf{Link}} & {\ul \textbf{NLOF}} & {\ul \textbf{Link}} & {\ul \textbf{NLOF}} & {\ul \textbf{Link}} & {\ul \textbf{NLOF}} \\
('128.163.217.2', 'UKS') & 0 & ('200.17.30.4', 'BS') & 0.383966245 & \textbf{('128.163.217.2', 'UKS')} & \textbf{0.108504399} \\
('128.163.217.3', 'UKS') & 0 & \textbf{('BS', 'R4')} & 0.330143541 & \textbf{('129.108.40.2', 'US1')} & \textbf{0.084057971} \\
('128.163.217.4', 'UKS') & 0 & ('R1', 'R4') & 0.330143541 & \textbf{('129.108.42.4', 'US3')} & \textbf{0.043147208} \\
('129.108.40.2', 'US1') & 0 & ('R1', 'R2') & 0.179347826 & ('US3', 'R2') & 0.032704403 \\
('129.108.40.3', 'US1') & 0 & ('129.108.40.3', 'US1') & 0.147058824 & ('UKS', 'R3') & 0.024746193 \\ \bottomrule
\end{tabular}
\end{table*}

\section{Conclusion}
\label{sec:conclusion}
By using multiple simulations in the NS-3 environment we have shown that it is possible to detect and localize soft-failures 
in a network using the Network Link Outlier Factor (NLOF). The results in Tables \ref{Tab:Topo1Results} and \ref{Tab:Topo2Results} 
show that the links with failures have the highest NLOF score which indicates where a fault in the network likely is. Using a new 
clustering technique, named TPCluster, we are able to provide a context for the performance of each individual flow. Our simulation 
experiments show that TPCluster yields meaningful clusters for identifying faults using outlier detection techniques. For future 
work we plan on studying the thresholds on NLOF scores for declaring a link failure. We also plan to expand the types of 
soft-failures we can detect.

\section{Acknowledgements}
This material is based upon work supported by both the U.S. Army Research Laboratory (USARL) under Cooperative Agreement 
W911NF-18-2-0287 and the National Science Foundation under Grant No. OAC-1450997.



\bibliographystyle{IEEEtran}
\bibliography{NLOF.bib}

\begin{thebibliography}{10}
\providecommand{\url}[1]{#1}
\csname url@samestyle\endcsname
\providecommand{\newblock}{\relax}
\providecommand{\bibinfo}[2]{#2}
\providecommand{\BIBentrySTDinterwordspacing}{\spaceskip=0pt\relax}
\providecommand{\BIBentryALTinterwordstretchfactor}{4}
\providecommand{\BIBentryALTinterwordspacing}{\spaceskip=\fontdimen2\font plus
\BIBentryALTinterwordstretchfactor\fontdimen3\font minus
  \fontdimen4\font\relax}
\providecommand{\BIBforeignlanguage}[2]{{%
\expandafter\ifx\csname l@#1\endcsname\relax
\typeout{** WARNING: IEEEtran.bst: No hyphenation pattern has been}%
\typeout{** loaded for the language `#1'. Using the pattern for}%
\typeout{** the default language instead.}%
\else
\language=\csname l@#1\endcsname
\fi
#2}}
\providecommand{\BIBdecl}{\relax}
\BIBdecl

\bibitem{DS0516}
A.~{Dusia} and A.~S. {Sethi}, ``Recent advances in fault localization in
  computer networks,'' \emph{IEEE Communications Surveys Tutorials}, vol.~18,
  no.~4, pp. 3030--3051, May 2016.

\bibitem{SS0704}
M.~Steinder and A.~S. Sethi, ``A survey of fault localization techniques in
  computer networks,'' \emph{Science of computer programming}, vol.~53, no.~2,
  pp. 165--194, July 2004.

\bibitem{MMHTOM0515}
M.~{Mukamoto}, T.~{Matsuda}, S.~{Hara}, K.~{Takizawa}, F.~{Ono}, and
  R.~{Miura}, ``Adaptive boolean network tomography for link failure
  detection,'' in \emph{2015 IFIP/IEEE International Symposium on Integrated
  Network Management (IM)}, May 2015, pp. 646--651.

\bibitem{D1206}
N.~{Duffield}, ``Network tomography of binary network performance
  characteristics,'' \emph{IEEE Transactions on Information Theory}, vol.~52,
  no.~12, pp. 5373--5388, Dec 2006.

\bibitem{CW1016}
M.~X. {Cheng} and W.~B. {Wu}, ``Data analytics for fault localization in
  complex networks,'' \emph{IEEE Internet of Things Journal}, vol.~3, no.~5,
  pp. 701--708, Oct 2016.

\bibitem{NSL0308}
M.~Natu, A.~S. Sethi, and E.~L. Lloyd, ``Efficient probe selection algorithms
  for fault diagnosis,'' \emph{Telecommunication Systems}, vol.~37, no. 1-3,
  pp. 109--125, March 2008.

\bibitem{CQMQB0310}
L.~{Cheng}, X.~{Qiu}, L.~{Meng}, Y.~{Qiao}, and R.~{Boutaba}, ``Efficient
  active probing for fault diagnosis in large scale and noisy networks,'' in
  \emph{2010 Proceedings IEEE INFOCOM}, March 2010, pp. 1--9.

\bibitem{NS1107}
M.~{Natu} and A.~S. {Sethi}, ``Probe station placement for fault diagnosis,''
  in \emph{IEEE GLOBECOM 2007 - IEEE Global Telecommunications Conference}, Nov
  2007, pp. 113--117.

\bibitem{MTG0818}
S.~{Madapuzi Srinivasan}, T.~{Truong-Huu}, and M.~{Gurusamy}, ``Te-based
  machine learning techniques for link fault localization in complex
  networks,'' in \emph{2018 IEEE 6th International Conference on Future
  Internet of Things and Cloud (FiCloud)}, Aug 2018, pp. 25--32.

\bibitem{VSRCCLPCCYV0118}
A.~Vela, B.~Shariati, M.~Ruiz, F.~Cugini, A.~Castro, H.~Lu, R.~Proietti,
  J.~Comellas, P.~Castoldi, S.-J.~B. Yoo \emph{et~al.}, ``Soft failure
  localization during commissioning testing and lightpath operation,''
  \emph{Journal of Optical Communications and Networking}, vol.~10, no.~1, pp.
  A27--A36, Jan 2018.

\bibitem{SMCT0318}
S.~{Shahkarami}, F.~{Musumeci}, F.~{Cugini}, and M.~{Tornatore},
  ``Machine-learning-based soft-failure detection and identification in optical
  networks,'' in \emph{2018 Optical Fiber Communications Conference and
  Exposition (OFC)}, March 2018, pp. 1--3.

\bibitem{TAB0505}
Y.~{Tang}, E.~S. {Al-Shaer}, and R.~{Boutaba}, ``Active integrated fault
  localization in communication networks,'' in \emph{2005 9th IFIP/IEEE
  International Symposium on Integrated Network Management, 2005. IM 2005.},
  May 2005, pp. 543--556.

\bibitem{TAB0308}
Y.~{Tang}, E.~{Al-Shaer}, and R.~{Boutaba}, ``Efficient fault diagnosis using
  incremental alarm correlation and active investigation for internet and
  overlay networks,'' \emph{IEEE Transactions on Network and Service
  Management}, vol.~5, no.~1, pp. 36--49, March 2008.

\bibitem{AAK0914}
N.~L. Van~Adrichem, B.~J. Van~Asten, F.~A. Kuipers \emph{et~al.}, ``Fast
  recovery in software-defined networks.'' \emph{EWSDN}, vol.~14, pp. 61--66,
  September 2014.

\bibitem{SSCPD1011}
D.~Staessens, S.~Sharma, D.~Colle, M.~Pickavet, and P.~Demeester, ``Software
  defined networking: Meeting carrier grade requirements,'' in \emph{18th IEEE
  Workshop on Local and Metropolitan Area Networks (LANMAN)}.\hskip 1em plus
  0.5em minus 0.4em\relax IEEE, October 2011, pp. 1--6.

\bibitem{MABPPRST0408}
N.~McKeown, T.~Anderson, H.~Balakrishnan, G.~Parulkar, L.~Peterson, J.~Rexford,
  S.~Shenker, and J.~Turner, ``Openflow: enabling innovation in campus
  networks,'' \emph{ACM SIGCOMM Computer Communication Review}, vol.~38, no.~2,
  pp. 69--74, March 2008.

\bibitem{PPTI0316}
F.~Pakzad, M.~Portmann, W.~L. Tan, and J.~Indulska, ``Efficient topology
  discovery in openflow-based software defined networks,'' \emph{Computer
  Communications}, vol.~77, pp. 52--61, March 2016.

\bibitem{EKSX0896}
M.~Ester, H.-P. Kriegel, J.~Sander, X.~Xu \emph{et~al.}, ``A density-based
  algorithm for discovering clusters in large spatial databases with noise.''
  in \emph{Kdd}, vol.~96, no.~34, August 1996, pp. 226--231.

\end{thebibliography}
%

\end{document}